 \documentstyle[12pt, a4]{article}
 \begin{document}
 \author{B. G\"{o}n\"{u}l, O. \"{O}zer, M. Ko\c{c}ak
 \and Department of Engineering Physics,Faculty of Engineering,\and
 University of Gaziantep, 27310 Gaziantep-T\"{u}rkiye}
 \title{Unified treatment of screening Coulomb and anharmonic
 oscillator potentials in arbitrary dimensions}
 \maketitle
 \begin{abstract}
 A mapping is obtained relating radial screened Coulomb systems
 with low screening parameters to radial anharmonic oscillators
 in $N$-dimensional space. Using the formalism of
 supersymmetric quantum mechanics, it is
 shown that exact solutions of these potentials exist when the
 parameters satisfy certain constraints.
 \end{abstract}

 Pacs Numbers: 03.65.Ca, 03.65.Ge, 11.30.Na

 \section{Introduction}
 Since the appearance of quantum mechanics there has been continual
 interest in models for which the corresponding Schr\"{o}dinger
 equation is exactly solvable. Solvable potential problems have
 played a dual role in physics. First, they represented useful aids
 in modeling realistic physical problems, and second, they offered
 an interesting field of investigation in their own right. Related
 to this latter area, the concept of solvability has changed to
 some extent in recent years. With regards to solvability of the
 Schr\"{o}dinger equation there are three interesting classes of
 the potentials.

 The first class is the exactly solvable potentials allowing
 to obtain in explicit form all energy levels and corresponding
 wave functions. The hydrogen atom and harmonic oscillator
 are the best-known examples of this type. The second class
 is the so-called quasi-exactly solvable (QES) potentials
 for which a finite number of eigenstates of corresponding
 Hamiltonian can be found exactly in explicit form. The
 first examples of such potentials were given in \cite{singh}.
 Subsequently several methods for generating partially
 solvable potentials were worked out, as a result many
 QES solvable potentials were found \cite{turbiner}. The third class
 is the conditionally-exactly solvable potentials \cite{dutra} for
 which the eigenvalue problem for the corresponding
 Hamiltonian is exactly solvable only when the parameters
 of the potential obey certain conditions.

 Although modern computational facilities permit the construction
 of solutions for any well-behaved potentials to any degree of
 accuracy, there remains continuing interest in exact solutions for
 a wider range of potentials. In connection with this, the
 technique of changing the independent coordinate has always been a
 useful tool in the solution of the Schr\"{o}dinger equation. For
 one thing, this allows something of a systematic approach,
 enabling one to recognize the equivalence of superficially
 unrelated quantum mechanical problems. For example, solvable
 Natanzon \cite{natanzon} potentials in nonrelativistic quantum
 mechanics are known to group into two disjoint classes depending
 on whether the Schr\"{o}dinger equation can be reduced to a
 hypergeometric or a confluent hypergeometric equation. All the
 potentials within each class are connected via point canonical
 transformations. Gangopadhyaya and his co-workers \cite{gango}
 established a connection also between the two classes with
 appropriate limiting procedures and redefinition of parameters,
 thereby inter-relating all known solvable potentials. In order for
 the Schr\"{o}dinger equation to be mapped into another
 Schr\"{o}dinger equation, there are severe restrictions on the
 nature of the coordinate transformation. Coordinate
 transformations which satisfy these restrictions give rise to new
 solvable problems. When the relationship between coordinates is
 implicit, then the new solution are only implicitly determined,
 while if the relationship is explicit then the newly found
 solvable potentials are also shape invariant\cite{gango,junker}.
 In a more specific special application of these ideas, Kostelecky
 et al. \cite{kostelecky1} were able to relate, using an explicit
 coordinate transformation, the Coulomb problem in $N$-dimensions
 with the $N$-dimensional harmonic oscillator. Other explicit
 applications of the coordinate transformation idea can be found in
 the review articles of Haymaker and Rau \cite{haymaker}.

 Moreover, recently an anti-isospectral transformation called also
 as duality transformation was introduced \cite{krajewska}. This
 transformation relates the energy levels and wave functions of two
 QES potentials. Many recent papers [10-12, and the references
 therein] have addressed this subject of coordinate transformation
 placing a particular emphasis on QES power-law potentials, which
 is also the subject of the present work in some extent. The
 generalization to higher dimensions of one-dimensional QES
 potentials was discussed in a recent paper \cite{mavromatis} and a
 few specific $N$-dimensional solutions were listed. In that work,
 applying a suitable transformation, these potentials were shown to
 be related to the isotropic oscillator plus Coulomb potential
 system and some normalized isolated solutions for this system were
 obtained.

 The importance in the study of QES potentials, apart
 from intrinsic academic interest, rests on the
 possibility of using their solutions to test the
 quality of numerical methods and in the possible
 existence of real physical systems that they could
 represent. For instance, anharmonic oscillators and
 screening Coulomb (or Yukawa) potentials represent
 simplified models of many situations found  in
 different field of physics. These problems have been
 studied for years and a general solution has not yet been found.

 The problem of quantum anharmonic oscillators has been the
 subject of much discussion for decades, both from an analytical
 and a numerical point of view, because of its important
 applications in quantum-field theory \cite{itzykson}, molecular
 physics \cite{reid}, and solid-state and statistical physics
 \cite{kittle,pathria}. Various methods have been used
 successfully for the one-dimensional anharmonic oscillators with various types of
 anharmonicities. Relatively less attention has been given to the
 anharmonic oscillators in higher-dimensional space because of the
 presence of angular-momentum states that make the problem more
 complicated. The recent works \cite{chaudhuri,morales} have shown
 that there are many interesting features of the anharmonic
 oscillators and the perturbed Coulomb problems in
 higher-dimensional space, and discussed the explicit dependence
 of these two potentials.

 Using the spirit of the works in Refs. \cite{chaudhuri,morales},
 we show the mappings between screened Coulomb potentials (with low
 screening parameters) and
 anharmonic oscillator potentials in $N$-dimensional space, which
 have not been previously linked. The connection between these
 potentials are also checked numerically by the use of the
 Lagrange-mesh calculation technique \cite{baye,hesse}. Next we
 study the $N$-dimensional screened Coulomb problem and higher
 order anharmonic oscillators within the framework of
 supersymmetric quantum mechanics (SSQM) \cite{cooper} and have
 shown that SSQM yields exact solutions for a single state only.

 In the following section, we outline the mapping procedure used
 through the article and give the applications. We also discuss in
 the same section exact supersymmetric treatments of the ground
 state solutions for the initial and transformed potentials.
 Analysis and discussion of the results obtained are given in
 section 3. Section 4 involves concluding remarks.

 \section{Mappings between the two distinct systems}
 The radial Schr\"{o}dinger equation for a spherically symmetric
 potential in $N$-dimensional space

 \begin{equation}
 -\frac{1}{2}\left[\frac{d^{2}R}{dr^{2}}+\frac{N-1}{r}\frac{dR}{dr}\right]+\frac{%
 \ell(\ell+N-2)}{2r^{2}}R =[E-V(r)]R
 \end{equation}
 is transformed to
 \begin{equation}
 -\frac{d^{2}\Psi}{dr^{2}}+\left[\frac{(M-1)(M-3)}{4r^{2}}+2V(r)\right]\Psi=2E\Psi
 \end{equation}
 where $\Psi(r)=r^{(N-1)/2}R(r)$, the reduced radial wave function
 , and $M=N+2\ell$. Note that the solutions for a particular
 central potential are the same as long as $M$ remains unaltered.
 For instance, the $s$-wave eigensolutions  and eigenvalues in
 four-dimensional space are identical to the $p$-wave solutions in
 two-dimensions.

 We substitute $r=\alpha\rho^{2}/2$ and $R=F(\rho)/\rho^{\lambda}$,
 suggested by the known transformations between the Coulomb and
 harmonic oscillator problems \cite{kostelecky1,bergmann} and used
 by \cite{chaudhuri,morales} to show the mappings between
 unperturbed Coulomb and anharmonic oscillator problems, and
 transform Eq.(1) to another Schr\"{o}dinger-like equation in
 $N^{\prime}=2N-2-2\lambda$ dimensional space with angular momentum
 $L=2\ell+\lambda$,
 \begin{equation}
 -\frac{1}{2}\left[\frac{d^{2}F}{d\rho^{2}}+\frac{N^{\prime}-1}{\rho}\frac{dF}{%
 d\rho}\right]+\frac{L(L+N^{\prime}-2)}{2\rho^{2}}F
 =[\hat{E}-\hat{V}(\rho)]F
 \end{equation}
 where
 \begin{equation}
 \hat{E}-\hat{V}(\rho)=E\alpha^{2}\rho^{2}-\alpha^{2}\rho^{2}V(\alpha%
 \rho^{2}/2)
 \end{equation}
 and $\alpha$ is a parameter to be adjusted properly. Thus, by this
 transformation, the $N$ -dimensional radial wave Schr\"{o}dinger
 equation with angular momentum $\ell $ can be transformed to a
 $N^{\prime}=2N-2-2\lambda$ dimensional equation with angular
 momentum $L=2\ell+\lambda$. The significance of the mapping
 parameter $\lambda$ is discussed below.

 A student of introductory quantum mechanics often learns that the
 Schr\"{o}dinger equation can be solved numerically for all angular
 momenta for the screened Coulomb and anharmonic oscillator
 problems. Less frequently, the student is made aware of the
 relation between these two problems, which are linked by a simple
 change of the independent variable discussed in detail through
 this section. Under this transformation, energies and coupling
 constants trade places, and orbital angular momenta are rescaled.
 Thus, we have shown here that there is really only one quantum
 mechanical problem, not two.

 The static screened Coulomb potential
 \begin{equation}
 V_{SC}(r)=-e^{2}\exp(-\delta r)/r
 \end{equation}
 where $\delta$ is a screening parameter, is known to describe
 adequately the effective interaction in many-body environment of a
 variety of fields such as atomic, nuclear, solid-state and plasma
 physics. In nuclear physics it goes under the name of the Yukawa
 potential (with $e^{2}$ replaced by another coupling constant),
 and in plasma physics it is commonly known as the Debye-H\"{u}ckel
 potential. Eq. (5) also describes the potential of an impurity in
 a metal and in a semiconductor \cite{weisbuch}. Since the
 Schr\"{o}dinger equation for such potentials does not admit exact
 analytic solutions, various approximate methods [23, 24, and the
 references therein], both analytic and numerical, have been
 developed.

 Considering the recent interest in various power-law potentials in
 the literature, we work through the article within the frame of
 low screening parameter. In this case, the screened Coulomb
 potential appears
 \begin{eqnarray}
 V_{SC}(r)&=&-e^{2}~\frac{\exp(-\delta r)}{r}\cong
 -\frac{e^{2}}{r}+e^{2}\delta-
 \frac{e^{2}\delta^{2}}{2}r+\frac{e^{2}\delta^{3}}{6}r^{^2}-
 \frac{e^{2}\delta^{4}}{24}r^{3}+
 \frac{e^{2}\delta^{5}}{120}r^{4}%
 \nonumber
 \end{eqnarray}
 \begin{eqnarray}
 &=&\frac{A_{1}}{r}+A_{2}+A_{3}r+A_{4}r^{2}+A_{5}r^{3}+A_{6}r^{4}
 \end{eqnarray}
 in the form. The literature is rich with examples of particular
 solutions for such power-law potentials employed in different
 fields of physics, for recent applications see Refs.
 \cite{znojil,alberg}. At this stage one may wonder why the series
 expansion is truncated at a lower order. This can be understood as
 follows. It is widely appreciated that convergence is not an
 important or even desirable property for series approximations in
 physical problems. Specifically, a slowly convergent approximation
 which requires many terms to achieve reasonable accuracy is much
 less valuable then a divergent series which gives accurate answers
 in a few terms. This is clearly the case for the screening Coulomb
 problem \cite{doren}. This also explains why leading orders of the
 $1/N$ expansion converge at a similar rate for the hydrogen atom, the
 screening Coulomb potential, and two-electron atom even though the
 last two of these series diverge eventually \cite{chatterjee}. In
 addition, the complexity of calculating higher order terms in the
 series for the corresponding transformed potential grows rapidly.
 Hence, if an accurate approximation cannot be achieved in a few
 terms, the present method may not be useful. However, in section 3
 we show that the present technique gives excellent estimates for
 the energy eigenvalues of both, the truncated screening Coulomb
 and anharmonic oscillator problems in higher dimensions.

 Though the mapping procedure introduced is valid for any bound state,
 throughout this work we are concerned ony with the ground state. We have two
 reasons for this restriction. The first reason is that the exact analytical ground state
 solutions for the potentials of interest are available, which provides a
 test for the reliability of the present technique. The second reason is that
 the present model works well for low lying states, which will be shown in section 3.
 Proceeding, therefore, with the choice of $\alpha^{2}=1/|E_{n=0}|$ in Eq. (4), the screened
 Coulomb problem in Eq. (6) is transformed to an anharmonic
 oscillator problem such that
 \begin{equation}
 \hat{V}(\rho)=\left(1+\frac{A_{2}}{|E_{n=0}|}\right)\rho^{2}+
 \frac{A_{3}}{2|E_{n=0}|^{3/2}}\rho^{4}+
 \frac{A_{4}}{4|E_{n=0}|^{2}}\rho^{6}+
 \frac{A_{5}}{8|E_{n=0}|^{5/2}}\rho^{8}+
 \frac{A_{6}}{16|E_{n=0}|^{3}}\rho^{10}
 \end{equation}
 with the eigenvalue
 \begin{equation}
 \hat{E}_{n=0}=\frac{-2A_{1}}{|E_{n=0}|^{1/2}}
 \end{equation}

 Thus, the system given by Eq. (6) in $N$-dimensional space is
 reduced to another system defined by Eq. (7) in
 $N^{\prime}=2N-2-2\lambda$ dimensional space. In other words, by
 changing the independent variable in the radial Schr\"{o}dinger
 equation, we have been able to demonstrate a close equivalence
 between the screened Coulomb potential and anharmonic oscillator
 potentials.

 For almost two decades, the study of higher order anharmonic
 potentials has been desirable to physicists and
 mathematicians in understanding a few newly discovered phenomena
 such as structural phase transitions \cite{khare}, polaron
 formation in solids \cite{emin}, and the concept of false vacuum
 in the field theory \cite{coleman}. Unfortunately, in these
 anharmonic potentials, not much work has been carried out on the
 potential like the one defined by (7) except the works in Refs.
 [32-34]. Our investigation in $N$-dimensional space, beyond
 showing an explicit connection between two distinct systems
 involving potentials of type (6) and (7), would also be helpful to
 the literature regarding the solutions of such potentials in
 arbitrary dimensions due to the recent wide interest in the
 lower-dimensional field theory.

 Eqs. (7) and (8) are the most important expressions in the present
 work. To test explicitly if these results are reliable, one needs
 to have an exact analytical solutions for the potentials of
 interest, which are found below within the frame of supersymmetric
 quantum mechanics.

 \subsection{Supersymmetric treatment for the ground state}
 Using the formalism of SSQM \cite{cooper} we set the
 superpotential
 \begin{equation}
 W(r)=\frac{a_{1}}{r}+a_{2}+a_{3}r+a_{4}r^{2},~a_{4}<0
 \end{equation}
 for the potential in (6) and identify $V_{+}(r)$ supersymmetric
 partner potential with the corresponding effective potential so
 that
 \begin{eqnarray}
 V_{+}(r)&=&W^{2}(r)+W^{\prime}(r)=\frac{2a_{1}a_{2}}{r}+
 [a_{2}^{2}+a_{3}(2a_{1}+1)]+2(a_{1}a_{4}+a_{4}+a_{2}a_{3})r
 \nonumber
 \end{eqnarray}
 \begin{eqnarray}
 +(2a_{2}a_{4}+a_{3}^{2})r^{2}
 +2a_{3}a_{4}r^{3}+a_{4}^{2}r^{4}+\frac{a_{1}(a_{1}-1)}{r^{2}}
 \nonumber
 \end{eqnarray}
 \begin{eqnarray}
 =\left(\frac{2A_{1}}{r}+2A_{2}+2A_{3}r+2A_{4}r^{2}+2A_{5}r^{3}+2A_{6}r^{4}\right)
 +\frac{(M-1)(M-3)}{4r^{2}}-2E_{n=0}
 \end{eqnarray}
 where $n=0,1,2,...$ is the radial quantum number. The relations
 between the parameters in (10) satisfy the supersymmetric
 definitions
 \begin{equation}
 a_{1}=\frac{M-1}{2},~a_{2}=\frac{2A_{1}}{M-1},~
 a_{3}=-\frac{A_{5}}{\sqrt{2A_{6}}},~a_{4}=-\sqrt{2A_{6}}
 \end{equation}
 Note that in order to retain the well-behaved solution at
 $r\rightarrow \infty$ we have chosen the negative sign in $a_{4}$.
 The potential in (6) admits the exact solutions
 \begin{equation}
 \Psi_{n=0}(r)= N_{0} r^{a_{1}}
 \exp\left(a_{2}r+\frac{a_{3}}{2}r^{2}+\frac{a_{4}}{3}r^{3}\right)
 \end{equation}
 where $N_{0}$ is the normalization constant, with the physically
 acceptable eigenvalues
 \begin{equation}
 E_{n=0}=A_{2}-\frac{1}{2}\left[\frac{4A_{1}^{2}}{(M-1)^{2}}-\frac{A_{5}}
 {\sqrt{2A_{6}}}M\right]
 \end{equation}
 under the constraints
 \begin{equation}
 A_{1}=-(M-1)\frac{8A_{6}A_{4}-2A_{5}^{2}}{16A_{6}\sqrt{2A_{6}}}~~,~~
 A_{3}=-\sqrt{2A_{6}}\left[\frac{(M+1)}{2}+\frac{A_{1}}{(M-1)}\frac{A_{5}}{A_{6}}\right]~,
 \end{equation}
 from which one arrives at the uniquely determined values of $M\approx 5$ and
 $\delta\approx 0.28$ in case of using atomic units in (6).
 The results obtained here are the generalization of the work in
 Ref. \cite{znojil}.

 For the  anharmonic oscillator potential in (7), we set the
 corresponding superpotential
 \begin{equation}
 W(\rho)=a\rho^{5}+b\rho^{3}+\frac{c}{\rho}+d\rho,~a<0,~d<0
 \end{equation}
 which leads to
 \begin{equation}
 \Psi_{n=0}(\rho)= C_{0} \rho^{c}
 \exp\left(\frac{a}{6}\rho^{6}+\frac{b}{4}\rho^{4}+\frac{d}{2}\rho^{2}\right)
 \end{equation}
 with $C_{0}$ being the corresponding normalization constant. Note that $\Psi(\rho)$ satisfies
 a differential equation analogous to Eq. (2) and is related to $F(\rho)$ in Eq. (3) as
 $\Psi(\rho)=\rho^{(N'-1)/2}~F(\rho)$ like the relation between $R(r)$ and $\Psi(r)$.
 Identifying $V_{+}(\rho)$ with the effective potential we arrive at
 an expression for the physically meaningful ground state
 eigenvalues of the anharmonic oscillator potential in arbitrary
 dimensions,
 \begin{equation}
 \hat{E}_{n=0}=-\frac{d}{2}(2c+1)=\frac{8A_{6}A_{4}-2A_{5}^{2}}{16A_{6}\sqrt{2A_{6}}}
 \frac{M'}{|E_{n=0}|^{1/2}}
 \end{equation}
 where $M'=N'+2L$, and the relations between the potential
 parameters satisfy the supersymmetric constraints
 \begin{equation}
 a=\pm\sqrt{\frac{A_{6}}{8}}\frac{1}{|E_{n=0}|^{3/2}},
 b=\frac{A_{5}}{8a}\frac{1}{|E_{n=0}|^{5/2}}, c=\frac{M'-1}{2},
 d=\frac{1}{2a}\left(\frac{A_{4}}{2|E_{n=0}|^{2}}-b^{2}\right)
 \end{equation}

 As we are dealing with a confined particle system, the negative
 values for $a$ and $d$ would of course be the right choice to
 ensure the well behaved nature of the wave function behaviour at
 infinity. Our results are in agreement with the
 literature existing for three-dimensions \cite{kaushal1,kaushal2}
 (in case $N'=3)$ and for two-dimensions \cite{dong} (in case
 $N'=2,L\rightarrow L-1/2$).

 In sum, we have shown that SSQM yields exact solutions for a single state only for
 the underlying quasi-exactly solvable potentials in higher dimensions with
 some constraints on the coupling constants. These constraints
 differ from each eigenvalue, and hence various solutions do not
 correspond to the same potential and are not orthogonal. We have
 not found these solutions discussed in the literature.

 \subsection{Significance of the mapping parameter}
 To show explicitly the physics behind the transformation described
 above, and to make clear the significance of the mapping parameter
 $\lambda$, we consider Eqs. (13) and (17) together within the same
 frame and arrive at
 \begin{equation}
 \hat{E}_{n=0}=-\frac{M'}{M-1}\frac{A_{1}}{|E_{n=0}|^{1/2}}
 \end{equation}

 To be consistent with Eq. (8) we must impose $0\leq\lambda\leq1$
 as an integer, such that
 \begin{equation}
 \frac{M'}{M-1}=\frac{2(N-1-\lambda)+2(2\ell+\lambda)}{N+2\ell-1}=2
 \end{equation}

 It is a general feature of this map that , in both cases
 ($\lambda=0,1$), the spectrum of the $N$-dimensional screened
 Coulomb problem is related to half the spectrum of the
 $N'$-dimension anharmonic oscillator for any even integer $N'$.
 The reader is referred to \cite{kostelecky1} for a comprehensive
 discussion of similar conformal mappings in the language of
 physics.

 It is worthwhile at this stage to note that recently Chaudhuri and
 Mondal \cite{chaudhuri} studied the relations between anharmonic
 oscillators and perturbed Coulomb potentials in higher dimensions
 but their results correspond only to the case when $\lambda=1$, in
 this case the three-dimensional perturbed Coulomb problem and the
 four-dimensional anharmonic oscillator cannot be related. However,
 by introducing an extra degree of freedom for the mapping
 parameter ($\lambda=0$) through our equations, we can reproduce
 the well-known results found in the literature in
 three-dimensions. With this exact correspondence we can check Eq.
 (8), using exact results for the screened Coulomb potential, and
 calculated numerical results for the anharmonic oscillator
 potential.

 \section{Results and Discussion}
 In this section numerical applications of the transformation
 presented in the previous section are given. Calculations to check
 the validity of the equations developed for the screened Coulomb
 and anharmonic potentials are also given here. Table I displays
 the exact eigenvalues of the screened Coulomb potential in three-,
 and five-dimensions obtained using the Lagrange-mesh calculation
 technique \cite{baye,hesse} for selected values of the potential
 parameters. Highly accurate Lagrange-mesh calculation results
 agree well with the best existing numerical and theoretical values
 obtained in three-dimensions \cite{dutt, lam}. Due to the
 constraint in the potential parameter $A_{1}$ expressed in Eq.
 (14), we are not able to show in the same table the corresponding
 exact energy values which can be calculated by Eq. (13). For the
 work of interest in this paper we set $A_{1}=-1$, consequently the
 adequate $\delta$-values satisfying the condition in Eq. (14) fall
 outside the scope of the presented work which has been performed
 for only low screening parameters.

 Further, our calculation results shown in Table I make clear that
 the eigenvalues of the five-dimensional screened Coulomb problem
 with any angular momentum quantum number $\ell$, for a particular
 $\delta$ value, are equal to the same system with $\ell+1$ in
 three-dimensions, due to $M=N+2\ell$ which remains unaltered for
 these states. The tabulated results support the work of Imbo and
 Sukhatme \cite{imbo} in which they formulated SSQM for spherically
 symmetric potentials in $N$ spatial dimensions and showed that
 the supersymmetric partner of a given potential can be effectively
 treated as being $N+2~(\ell\rightarrow\ell+1)$ dimensions. This
 fact was exploited in their calculations using the shifted $1/N$
 expansion.

 It is also noted that for very small values of the screening
 parameter, the screening Coulomb potential reduces to the Coulomb
 potential that is shape invariant having supersymmetric
 character. Therefore, the related supersymmetric partner
 potentials, such as $V_{\ell}$ and $V_{\ell+1}$, are expected to
 have the same spectra except the ground-state energy. This can
 easily be seen in Table I for the case of $\delta=0.001$ in both
 arbitrary dimensions. For instance, the supersymmetric partner of
 the $s$-orbital ($\ell=0$) spectrum of hydrogen is the $p$-orbital
 ($\ell=1$) spectrum of the same system.

 Finally, the exact calculated eigenvalues, by the use of Eq. (8),
 for the anharmonic oscillator in four-dimensions from the known
 exact results for the screened Coulomb problem in three-dimensions
 are displayed in Table II. These eigenvalues are checked by the
 Lagrange-mesh calculations and tabulated in the same table. For
 low lying states, the results obtained with the present technique
 agree well with the numerical calculations, but this agreement
 deteriorates quickly for higher lying states.

 \section{Concluding Remarks}
 The mapping problems in arbitrary dimensions have been the subject
 of several papers and have served to illustrate various aspects of
 quantum mechanics of considerable pedagogical value. As the
 objective of this presentation we have highlighted a different
 facet of these studies and established a very general connection
 between the screened Coulomb and anharmonic oscillator potentials
 in higher dimensional space through the application of a suitable
 transformation, the purpose being the emphasize the pedagogical
 value residing in this interrelationship between two of the most
 practical applications of quantum mechanics. The exact ground
 state solutions for the potentials considered are obtained
 analytically within the framework of supersymmetric quantum
 mechanics, which provides a testing ground for benchmark
 calculations. Although much work had been done in the literature
 on similar problems, an investigation as the one we have discussed
 in this paper was missing to our knowledge.

 \newpage\

 \begin{center}
 {\bf Acknowledgments}
 \end{center}

 We are grately indebted to Daniel Baye for his kindness in
 providing us the modified version of the computer code to perform
 Lagrange-mesh calculations. We are also grateful to the referees for suggesting
 several amendments.

 \begin{table}
 \caption{\label{TABLE I}
         {The first four eigenvalues of the screening Coulomb
         potential in Eq. (6) as a function of the screening
         parameter in atomic units.}}
 \vspace{5mm}
 \begin{center}
 \begin{tabular}{cccccc}
 \hline \hline
 \\
 \multicolumn{6}{c}{In three-dimensional space}
 \\
 \hline \\
 $\delta$&$\ell$&n=0&n=1&n=2&n=3\\ \hline 0.001&0&-0.499
 000&-0.124 003&-0.054 562&-0.030 262\\
      &1&-0.124 002&-0.054 561&-0.030 261&-0.019 018\\
      &2&-0.054 561&-0.030 260&-0.019 017&-0.012 914\\
 0.005&0&-0.495 019&-0.120 074&-0.050 720&-0.026 537\\
      &1&-0.120 062&-0.050 708&-0.026 526&-0.015 428\\
      &2&-0.050 684&-0.026 503&-0.015 406&-0.009 474\\
 0.010&0&-0.490 075&-0.115 293&-0.046 199&-0.022 356\\
      &1&-0.115 245&-0.046 153&-0.022 313&-0.011 622\\
      &2&-0.046 061&-0.022 228&-0.011 543&-0.006 070\\
 0.020&0&-0.480 296&-0.106 148&-0.038 020&-0.015 377\\
      &1&-0.105 963&-0.037 852&-0.015 232&-0.005 891\\
      &2&-0.037 515&-0.014 939&-0.005 653&-0.001 521\\
 0.025&0&-0.475 461&-0.101 776&-0.034 329&-0.012 495\\
      &1&-0.101 492&-0.034 079&-0.012 287&-0.003 770\\
      &2&-0.033 573&-0.011 865&-0.003 458& 0.000 253\\
 \\
 \hline \\
  \multicolumn{6}{c}{In five-dimensional space}
 \\
 \hline \\
 0.001&0&-0.124 002&-0.054 561&-0.030 261&-0.019 018\\
      &1&-0.054 561&-0.030 260&-0.019 017&-0.012 914\\
      &2&-0.030 259&-0.019 016&-0.012 912&-0.009 237\\
 0.005&0&-0.120 062&-0.050 708&-0.026 526&-0.015 428\\
      &1&-0.050 684&-0.026 503&-0.015 406&-0.009 474\\
      &2&-0.026 468&-0.015 373&-0.009 443&-0.005 961\\
 0.010&0&-0.115 245&-0.046 153&-0.022 313&-0.011 622\\
      &1&-0.046 061&-0.022 228&-0.011 543&-0.006 070\\
      &2&-0.022 099&-0.011 425&-0.005 965&-0.002 980\\
 0.020&0&-0.105 963&-0.037 852&-0.015 232&-0.005 891\\
      &1&-0.037 515&-0.014 939&-0.005 653&-0.001 521\\
      &2&-0.014 491&-0.005 286&-0.001 263& 0.000 885\\
 0.025&0&-0.101 492&-0.034 079&-0.012 287&-0.003 770\\
      &1&-0.033 573&-0.011 865&-0.003 458& 0.000 253\\
      &2&-0.011 216&-0.002 974& 0.000 524& 0.003 087\\
 \hline \hline
 \end{tabular}
 \end{center}
 \end{table}
 \newpage\
 \begin{table}[t]
 \caption{\label{TABLE II}
         { Ground-state eigenvalues of the anharmonic potential in Eq. (7)}}
 \vspace{5mm}
 \begin{center}
 \begin{tabular}{cccccc}
 \hline \hline
 \\
 \multicolumn{6}{c}{In four-dimensional space}
 \\
 \hline \\
 $\delta$&$\ell$&L&$|E_{n=0}|$&$\hat{E}_{n=0}$&$\hat{E}_{n=0}$
 \\
 &&&(taken from Table I)&Lagrange-mesh calculations&Exact value
 (Eq. 8)
 \\
 \hline \\
 0.001&0&0&0.499 000&2.831 259&2.831 259\\
      &1&2&0.124 002&5.679 579&5.679 573\\
      &2&4&0.054 561&8.562 285&8.562 268\\
 0.005&0&0&0.495 019&2.842 624&2.842 622\\
      &1&2&0.120 062&5.772 014&5.772 012\\
      &2&4&0.050 684&8.883 704&8.883 714\\

 0.010&0&0&0.490 075&2.856 927&2.856 924\\
      &1&2&0.115 245&5.891 401&5.891 406\\
      &2&4&0.046 061&9.318 882&9.318 871\\

 0.020&0&0&0.480 296&2.885 862&2.885 862\\
      &1&2&0.105 963&6.144 014&6.144 024\\
      &2&4&0.037 515&10.325 883&10.325 891\\

 0.025&0&0&0.475 461&2.900 499&2.900 498\\
      &1&2&0.101 492&6.277 884&6.277 896\\
      &2&4&0.033 573&10.915 282&10.915 281\\
 \hline \hline
 \end{tabular}
 \end{center}
 \end{table}
\end{document}